\documentclass[12pt, a4paper]{article}
\pdfoutput=1

\usepackage{cite}
\usepackage{amsmath,amssymb}
\usepackage{xcolor}
\usepackage{graphicx}
\usepackage[bookmarksopen,colorlinks=false]{hyperref}

\begin{document}

%%%%%%%%%%%%%%%%%%%%%%%%%%%%%%%%%%%%%%%%%%%%%%%%%%
\begin{titlepage}

\begin{center}

\hfill UT-17-31

\vskip .75in

{\Large \bf 
Explosive Axion Production from Saxion
}

\vskip .75in

{\large
Yohei Ema$^{(a)}$ and Kazunori Nakayama$^{(a,b)}$
}

\vskip 0.25in

$^{(a)}${\em Department of Physics, Faculty of Science,\\
The University of Tokyo,  Bunkyo-ku, Tokyo 113-0033, Japan}\\[.3em]
$^{(b)}${\em Kavli IPMU (WPI), UTIAS,\\
The University of Tokyo,  Kashiwa, Chiba 277-8583, Japan}

\end{center}
\vskip .5in

\begin{abstract}

The dynamics of saxion in a supersymmetric axion model and its effect on the axion production is studied in detail.
We find that the axion production is very efficient 
when the saxion oscillation amplitude is much larger than the Peccei-Quinn scale,
due to a spike-like behavior of the effective axion mass.
We also consider the axino production and several cosmological consequences.
The possibility of detection of gravitational waves 
from the non-linear dynamics of the saxion and axion is discussed.

\end{abstract}

\end{titlepage}

%\tableofcontents

\renewcommand{\thepage}{\arabic{page}}
\setcounter{page}{1}
\renewcommand{\thefootnote}{\#\arabic{footnote}}
\setcounter{footnote}{0}
%%%%%%%%%%%%%%%%%%%%%%%%%%%%%%%%%%%%%%%%%%%%%%%%%%

\newpage

%%%%%%%%%%%%%%%%%%%%%%%%%%%%%%%%%%%%%%%%%%%%%%%%%%
\section{Introduction}
\label{sec:}
\setcounter{equation}{0}
%%%%%%%%%%%%%%%%%%%%%%%%%%%%%%%%%%%%%%%%%%%%%%%%%%

Supersymmetric (SUSY) axion models combine the idea of SUSY 
and the Peccei-Quinn (PQ) mechanism~\cite{Peccei:1977hh} 
to solve both the hierarchy problem and the strong CP problem.
SUSY axion models are also interesting from the cosmological point of view,
mainly due to the existence of (relatively) light particles, the saxion and axino,
which are the scalar and fermionic partners of the axion, respectively~\cite{Kim:1983ia,Rajagopal:1990yx,Kim:1992eu,Lyth:1993zw,Chun:1995hc,Choi:1996vz,Kasuya:1996ns,Asaka:1998ns,Covi:2001nw,Brandenburg:2004du,Strumia:2010aa,Kawasaki:2007mk,Kawasaki:2008jc,Kawasaki:2010gv,Chun:2011zd,Bae:2011jb,Bae:2011iw,Jeong:2012np,Nakayama:2012zc,Moroi:2012vu,Choi:2012zna,Kawasaki:2013ae,Moroi:2013tea,Bae:2013hma,Co:2017orl}.

The cosmological saxion dynamics is highly non-trivial and its cosmological effects 
are still not fully understood despite its importance.
In a simplified treatment, the saxion is assumed to begin a coherent oscillation at some epoch 
and eventually it decays into an axion pair as well as the minimal SUSY standard model (MSSM) sector.
The produced axions compose dark radiation which is constrained from cosmological observations.
In such a scenario, it is often assumed that the saxion perturbatively decays into the axion pair.
It is still non-trivial, however, whether the axion production can be treated perturbatively especially 
when the saxion field value significantly deviates from the potential minimum.

In this paper we revisit the saxion dynamics in a model described by the superpotential~\eqref{super} 
shown below to correctly estimate the axion production rate.
Ref.~\cite{Kasuya:1996ns} partly addressed this issue and found that there is a parametric resonant enhancement of the axion
when the saxion amplitude is large enough.
We analyze the dynamics in more detail and identify the origin of the resonant axion production.
We find a ``spike''-like feature of the effective axion mass, 
which originates purely from multi-field property of the dynamics.
Due to the spike, the axion production rate is much more enhanced 
compared to the naive perturbative calculation of the decay rate.
Actually the axion production is so explosive that the whole saxion-axion system becomes completely non-linear
after only a few saxion oscillations.
It may drastically change the conventional picture of the saxion oscillation and the resulting particle production processes.

We first discuss the saxion dynamics in Sec.~\ref{sec:saxion}
in terms of the canonically normalized fields.
We estimate the axion and axino production rate in Sec.~\ref{sec:prod}.
Several cosmological implications are also mentioned.
We conclude with some discussion including implications for gravitational wave observation 
in Sec.~\ref{sec:dis}.

%%%%%%%%%%%%%%%%%%%%%%%%%%%%%%%%%%%%%%%%%%%%%%%%%%
\section{Saxion dynamics}
\label{sec:saxion}
\setcounter{equation}{0}
%%%%%%%%%%%%%%%%%%%%%%%%%%%%%%%%%%%%%%%%%%%%%%%%%%

%%%%%%%%%%%%%%%%%%%%%%%%%%%%%%%%%%%%%%%%%%%%%%%%%%
\subsection{Model}
%%%%%%%%%%%%%%%%%%%%%%%%%%%%%%%%%%%%%%%%%%%%%%%%%%

We consider the axion model in which the K\"ahler and superpotential are given by
\begin{align}
	K &= |\phi|^2 + |\bar\phi|^2 +|X|^2,\\
	W& = \lambda X(\phi\bar\phi - f^2) + W_0,  \label{super}
\end{align}
where $\phi$ and $\bar\phi$ are PQ superfields with PQ charges $+1$ and $-1$ respectively, 
and $X$ is a PQ-singlet superfield.
The constants $\lambda$, $f$ and $W_0$ are taken to be real and positive without loss of generality.
The $R$-symmetry under which only $X$ has a charge $+2$ ensures this type of superpotential.
Including the SUSY breaking effects, the scalar potential is given as
\begin{align}
	V = m_\phi^2|\phi|^2 + m_{\bar\phi}^2|\bar\phi|^2 + \lambda^2\left(\left|\phi\bar\phi - f^2\right|^2+|X|^2(|\phi|^2+|\bar\phi|^2)\right) + 2\lambda m_{3/2}f^2(X+X^\dagger),
\end{align}
where $m_\phi$ and $m_{\bar\phi}$ are soft SUSY breaking masses and $m_{3/2} = W_0/M_P^2$ denotes the gravitino mass with 
$M_P$ being the reduced Planck scale.
Below we assume $f\gg m_\phi, m_{\bar\phi}, m_{3/2}$.
The potential minimum is
\begin{align}
	&\left<X\right> = -\frac{2m_{3/2}f^2/\lambda}{\left<|\phi|\right>^2+\left<|\bar\phi|\right>^2},\\
	&\left<|\phi|\right> = f \left( \frac{m_{\bar\phi}^2 + \lambda^2\left< X\right>^2}{m_{\phi}^2 + \lambda^2\left< X\right>^2}\right)^{1/4} 
	+ \mathcal O\left(\frac{m_\phi^2}{f}\right),\\
	&\left<|\bar\phi|\right> = f \left( \frac{m_{\phi}^2 + \lambda^2\left< X\right>^2}{m_{\bar\phi}^2 + \lambda^2\left< X\right>^2}\right)^{1/4}
	+ \mathcal O\left(\frac{m_{\bar\phi}^2}{f}\right).
\end{align}
Below we redefine $m_\phi^2 +  \lambda^2\left< X\right>^2$ and $m_{\bar\phi}^2 + \lambda^2\left< X\right>^2$
as $m_\phi^2$ and $m_{\bar\phi}^2$, respectively.

In the SUSY limit $m_\phi, m_{\bar\phi}\to 0$, there are two massless modes, called axion and saxion,
corresponding to the flat direction $\phi\bar\phi = f^2$.
The saxion obtains a mass of the order of the soft SUSY breaking, but still it is much lighter than the PQ scale $\sim f$.\footnote{
	The PQ scale $f_a$ is given by $f_a^2 = (\varphi^2 + \bar\varphi^2)/N_{\rm DW}^2$
	with $N_{\rm DW}$ being the domain wall number, which depends on the PQ charge assignments on the MSSM sector.
}
Thus the saxion dynamics can have significant impact on cosmology.
In particular, we are interested in the case that the saxion has an initial value as large as $M_P$.\footnote{
	Actually, if the PQ field obtains a negative Hubble-induced mass squared through the coupling to the inflaton,
	the saxion gets a VEV of $\sim M_P$ during inflation.
}
We reconsider the saxion dynamics in such a case, paying particular attention to its effect on the axion production.

%%%%%%%%%%%%%%%%%%%%%%%%%%%%%%%%%%%%%%%%%%%%%%%%%%
\subsection{Equation of motion}
%%%%%%%%%%%%%%%%%%%%%%%%%%%%%%%%%%%%%%%%%%%%%%%%%%

Let us write the PQ fields as
\begin{align}
	\phi = \frac{\varphi}{\sqrt 2}e^{ia_\phi/\varphi},~~~~~~\bar\phi = \frac{\bar\varphi}{\sqrt 2}e^{ia_{\bar\phi}/\bar\varphi}.
\end{align}
The dynamics is almost constrained to the flat direction $\phi\bar\phi=f^2$ since
the scalar degrees orthogonal to it are heavy enough.
This constraint is written as
\begin{align}
	\varphi\bar\varphi = 2f^2,~~~~~~
	\frac{a_\phi}{\varphi} + \frac{a_{\bar\phi}}{\bar\varphi} = 0.
\end{align}
Using this, we can replace $\bar\varphi$ and $a_{\bar\phi}$ in terms of $\varphi$ and $a_\phi$
which correspond to the saxion and axion, respectively.
Thus we obtain the kinetic term for the saxion and axion as
\begin{align}
	\mathcal L_K&=-|\partial\phi|^2-|\partial\bar\phi|^2 \\
	&=-\frac{1}{2}F^2(\varphi)\left[\left(\partial\varphi)^2+(\partial a_\phi\right)^2 
	+\frac{\partial^2\varphi}{\varphi}a_\phi^2\right] 
	-F F_\varphi\frac{(\partial\varphi)^2}{\varphi}a_\phi^2,
\end{align}
where
\begin{align}
	F(\varphi) \equiv  \sqrt{1+\frac{4f^4}{\varphi^4}},~~~~F_\varphi =-\frac{8}{F}\frac{f^4}{\varphi^5}.
\end{align}
The scalar potential is
\begin{align}
	V \simeq m_\phi^2|\phi|^2 + m_{\bar\phi}^2|\bar\phi|^2 
	= \frac{m_\phi^2}{2}\varphi^2F^2(\varphi)+\frac{2\Delta m^2 f^4}{\varphi^2},
\end{align}
where $\Delta m^2\equiv m^2_{\bar\phi}-m^2_{\phi}$.
The equation of motion of $\varphi$ reads
\begin{align}
	\partial^2\varphi 
	&= m_\phi^2\varphi \left(1-\frac{4f^4}{\varphi^4}\right) F^{-2} -\frac{F_\varphi}{F}(\partial\varphi)^2-\frac{4\Delta m^2 f^4}{F^2\varphi^3},
	\label{eom_saxion}
\end{align}
as far as the backreaction from $a_\phi$ is neglected.
In the small amplitude limit $\varphi = \left<\varphi\right>+\delta\varphi$ $(|\delta\varphi| \ll f)$, 
it is simplified as
\begin{align}
	\partial^2\delta\varphi \simeq \frac{4m_\phi^2}{F^2}\delta\varphi=
	\frac{4m_\phi^2 m_{\bar\phi}^2}{m_\phi^2+m_{\bar\phi}^2}\delta\varphi \equiv \widetilde m^2_{\phi} \delta\varphi.
\end{align}
The equation of motion of $a_\phi$ reads
\begin{align}
	\partial^2 a_\phi + \frac{2F_\varphi}{F}\partial \varphi \,\partial a_\phi
	- \left(\frac{\partial^2 \varphi}{\varphi} + \frac{2F_\varphi}{F}\frac{\left(\partial \varphi\right)^2}{\varphi} \right)a_\phi
	= 0.
\end{align}
It may be useful to rewrite it in terms of $\theta\equiv a_\phi/\varphi$.
The kinetic term for $\theta$ is given as
\begin{align}
	\mathcal L_K = -\frac{1}{2}F^2(\varphi) \left[ (\partial\varphi)^2+\varphi^2(\partial\theta)^2 \right].
\end{align}
Then the equations of motion for $\theta$ is
\begin{align}
	\partial^2 \theta + 2\left(1 - \frac{4f^4}{\varphi^4}\right)\frac{\partial\varphi}{F^2\varphi}
	(\partial\theta)=0.  \label{eom_theta}
\end{align}
Note that it has a shift symmetry $\theta \rightarrow \theta + \mathrm{const}$.

%%%%%%%%%%%%%%%%%%%%%%%%%%%%%%%%%%%%%%%%%%%%%%%%%%
\subsection{Canonical saxion and axion}
%%%%%%%%%%%%%%%%%%%%%%%%%%%%%%%%%%%%%%%%%%%%%%%%%%

For later convenience, we rewrite the equations of motion in terms of the canonically normalized fields.
Let us define the canonical saxion $\widetilde\varphi$ and canonical axion $\widetilde a_\phi$ as
\begin{align}
	\widetilde\varphi \equiv \int_{\left<\varphi\right>}^\varphi d\varphi F(\varphi),~~~~~~\widetilde a_\phi\equiv a_\phi F(\varphi).
\end{align}
Note that
\begin{align}
	\widetilde\varphi \simeq  \begin{cases}
		\varphi & {\rm for} ~~~\varphi \gg \left<\varphi\right>,\\
		\displaystyle F \delta\varphi =\sqrt{\frac{m_\phi^2+m_{\bar\phi}^2}{m_{\bar\phi}^2}}\delta\varphi & {\rm for} ~~~\varphi \simeq \left<\varphi\right>,\\
		-2f^2/\varphi & {\rm for} ~~~\varphi \ll \left<\varphi\right>.
	\end{cases}
\end{align}
The equation of motion of $\widetilde\varphi$ is
\begin{align}
	\partial^2\widetilde\varphi &
	= \frac{m_\phi^2\varphi}{F} \left(1-\frac{4f^4}{\varphi^4}\right) - \frac{4\Delta m^2 f^4}{F\varphi^3}\\
	&\simeq\begin{cases}
		m_\phi^2 \widetilde\varphi & {\rm for} ~~~\varphi \gg \left<\varphi\right>,\\
		\widetilde m_\phi^2 \widetilde\varphi & {\rm for} ~~~\varphi \simeq \left<\varphi\right>,\\
		m_{\bar\phi}^2 \widetilde\varphi & {\rm for} ~~~\varphi \ll \left<\varphi\right>.
	\end{cases}
\end{align}
Thus the canonical saxion $\widetilde \varphi$ behaves almost as a harmonic oscillator 
with its initial value $\widetilde{\varphi}_\mathrm{ini}$.
The equation of motion of the canonical axion $\widetilde{a}_\phi$ is
\begin{align}
	\partial^2 \widetilde{a}_\phi = m_a^2 \widetilde{a}_\phi,
\end{align}
where
\begin{align}
	m_{a}^2 
	= \left(1+\frac{\varphi F_\varphi}{F}\right)\frac{\partial^2\varphi}{\varphi} 
	+ \left(\frac{2F_\varphi}{F \varphi}+\frac{F_{\varphi\varphi}}{F} \right)(\partial\varphi)^2.
\end{align}
Using the equation of motion of $\varphi$, it is rewritten as
\begin{align}
	m_{a}^2
	&= \left(1-\frac{4f^4}{\varphi^4}\right)\left[\left(1-\frac{4f^4}{\varphi^4}\right) \frac{m_\phi^2}{F^4} 
	- \frac{4\Delta m^2 f^4}{F^4\varphi^4}\right]
	- \frac{4F_\varphi}{F^3 \varphi}(\partial\varphi)^2. 
	\label{ma2}
\end{align}
It is approximated as
\begin{align}
	m_{a}^2 \simeq\begin{cases}
		m_\phi^2 & {\rm for} ~~~\varphi \gg \left<\varphi\right>,\\
		\displaystyle \frac{\widetilde m_\phi^2 \Delta m^2}{F^2 m_{\bar\phi}^2}\frac{\delta\varphi}{ \left<\varphi\right>} 
		+\frac{\widetilde m_\phi^4}{m_{\bar\phi}^2}\frac{\delta\varphi^2}{\left<\varphi\right>^2} - \frac{32}{F^4}\frac{f^4\dot\varphi^2}{\varphi^6}
		& {\rm for} ~~~\varphi \simeq \left<\varphi\right>,\\
		m_{\bar\phi}^2 & {\rm for} ~~~\varphi \ll \left<\varphi\right>. 
	\end{cases}
	\label{ma}
\end{align}
Thus the axion is massless only if the background is settled to the potential minimum ($\delta\varphi=0$ and $\dot\varphi=0$).
It obtains effective mass if the background deviates from the minimum and time-dependent.
Especially, when the saxion is oscillating with its amplitude much larger than $f$, 
the last term in the second line of (\ref{ma}) becomes important as we will see below.

%%%%%%%%%%%%%%%%%%%%%%%%%%%%%%%%%%%%%%%%%%%%%%%%%%
\section{Particle production rate}  \label{sec:prod}
%%%%%%%%%%%%%%%%%%%%%%%%%%%%%%%%%%%%%%%%%%%%%%%%%%

%%%%%%%%%%%%%%%%%%%%%%%%%%%%%%%%%%%%%%%%%%%%%%%%%%
\subsection{Axion}  \label{sec:axion}
%%%%%%%%%%%%%%%%%%%%%%%%%%%%%%%%%%%%%%%%%%%%%%%%%%

%%%%%%%%%%%%%%%%%%%%%%%%%%%%%%%%%%%%%%%%%%%%%%%%%%
\subsubsection{Small saxion amplitude regime}
%%%%%%%%%%%%%%%%%%%%%%%%%%%%%%%%%%%%%%%%%%%%%%%%%%

First let us consider the small saxion amplitude case: $\widetilde \varphi_{\rm amp} \ll f$, to compare with the conventional calculation
of the saxion decay rate.
In this case, the term proportional to $\dot\varphi^2$ is negligible
in the axion mass expression (second line of  (\ref{ma})).
Then we can extract the axion production rate from the $\delta\varphi$ dependence of the axion mass term.
Approximating $\delta\varphi$ as a harmonic oscillator, we obtain the saxion decay rate into the axion pair as
\begin{align}
	\Gamma_{\varphi\to a} \simeq {\rm max}\left[\frac{1}{32\pi}\left(\frac{\Delta m^2}{F^3 m_{\bar\phi}^2}\right)^2 \frac{\widetilde m_\phi^3}{\left<\varphi\right>^2},
	~~~\frac{1}{4\pi}\frac{\widetilde m_\phi^4}{F^4 m_\phi^4} \frac{\widetilde\varphi_{\rm amp}^2}{\left<\varphi\right>^2} \frac{\widetilde m_\phi^3}{\left<\varphi\right>^2}   \right].
	\label{Gamma_small}
\end{align}
The first term is understood as the usual perturbative decay rate of the saxion into the axion pair around the vacuum.
The second term may be regarded as annihilation of the saxion into the axion pair, 
which exists for finite saxion amplitude.
As is known, the perturbative decay rate vanishes in the limit $\Delta m^2=0$~\cite{Chun:1995hc}.
Even in such a case, there is a finite contribution to the axion production from the saxion annihilation represented by the second term, although it cannot lead to the complete saxion decay since it decreases faster than the Hubble rate.

However, the axion production is very efficient in the large saxion oscillation regime even in the case of $\Delta m^2=0$ as we will show below.
In the following we take $\Delta m^2=0$ to derive the lower bound on the axion dark radiation abundance.

%%%%%%%%%%%%%%%%%%%%%%%%%%%%%%%%%%%%%%%%%%%%%%%%%%
\subsubsection{Large saxion amplitude regime}
%%%%%%%%%%%%%%%%%%%%%%%%%%%%%%%%%%%%%%%%%%%%%%%%%%

\begin{figure}[t]
\begin{center}
\begin{tabular}{cc}
\includegraphics[scale=1.2]{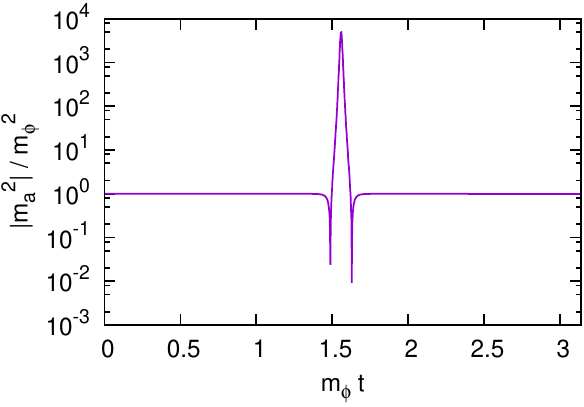}
\hspace{5mm}
\includegraphics[scale=1.2]{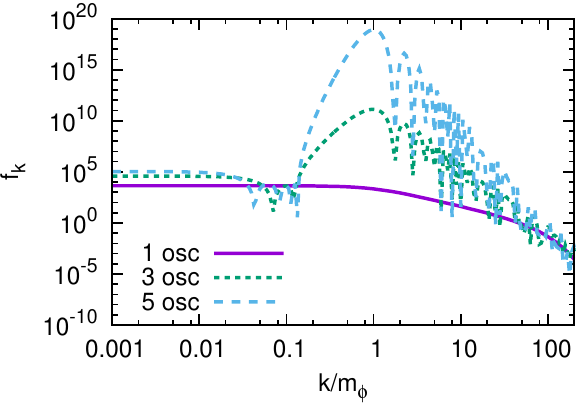}
\end{tabular}
\end{center}
\caption {
(Left) Time evolution of $|m_a^2(t)|$ for $\varphi_{\rm amp} = 100f$.
(Right) The phase space density of the axion $f_k$ as a function of wave number $k$ in units of $m_\phi$
for 1, 3 and 5 half saxion oscillation $(t=\pi/m_\phi, 3\pi/m_\phi, 5\pi/m_\phi)$ for $\varphi_{\rm amp} = 100f$.
Hence the spike scale is $m_{\rm sp} \sim 100 m_\phi$.
}
\label{fig:fk_axion}
\end{figure}

Let us consider a large saxion oscillation amplitude regime: $\widetilde \varphi_{\rm amp} \gg f$.\footnote{
	We express the amplitude of the saxion oscillation by $\widetilde\varphi_{\rm amp}$ and it is weakly time-dependent due to the Hubble expansion.
	It scales as $\widetilde\varphi_{\rm amp} \propto R(t)^{-3/2}$ with $R(t)$ being the cosmic scale factor. 
}
We assume $m_\phi \widetilde \varphi_{\rm amp} < f^2$ 
since otherwise the saxion dynamics is not confined to the flat direction
$\varphi\bar\varphi = 2f^2$ and the dynamics would be more chaotic, 
which may lead to the nonthermal PQ symmetry restoration.\footnote{
	This condition also guarantees that production of $X$ due to the saxion oscillation is negligible since
	$X$ has a mass of order $f$ even when the saxion crosses the potential minimum.
} The saxion oscillation is approximated as
\begin{align}
	\widetilde \varphi \simeq \widetilde \varphi_{\rm amp}\cos(m_\phi t).
\end{align}
Then we have
\begin{align}
	m_{a}^2 \simeq \begin{cases}
	\displaystyle m_\phi^2 & {\rm for}~~~\left\lvert \widetilde \varphi \right\rvert \gg f, \\
	\displaystyle -m_\phi^2\left[\frac{ \widetilde \varphi_{\rm amp}^2}{8f^2}\right] 
	& {\rm for}~~~\left\lvert \widetilde \varphi \right\rvert \ll f.
	\end{cases}
	\label{ma_spike}
\end{align}
Notice that the axion temporary becomes very massive when the saxion passes through the potential minimum $\widetilde\varphi \simeq 0$
where $m_{a}^2 \sim m_\phi^2 \widetilde \varphi_{\rm amp}^2 / f^2 \gg m_\phi^2$.
This is much different from the case of single PQ field models: in the single PQ field models,
the canonical axion mass is always zero when the saxion passes through the potential minimum.
A crucial difference between the single and multi PQ field model is that the definition of ``axion'' becomes time-dependent in the multi-field model.
Actually one easily recognizes that $\tilde a \simeq a_\phi$ for $\varphi \gg f$ and $\tilde a \simeq a_{\bar\phi}$ for $\varphi \ll f$.
The eigenstate suddenly changes around $\varphi \simeq f$ and the time scale for this sudden transition is
$\Delta t \sim \left(f / \dot\varphi\right)_{\varphi\simeq f} \sim f/(m_\phi \widetilde\varphi_{\rm amp})$.
This is the reason for the appearance of temporal large mass scale in the effective axion mass.
We call this as a ``spike''-like behavior of the effective axion mass due to the saxion dynamics.
The left panel of Fig.~\ref{fig:fk_axion} shows time evolution of $|m_a^2(t)|$ for $\varphi_{\rm amp} = 100f$
by numerically solving the equation of motion of the saxion (\ref{eom_saxion}) with $\Delta m^2 = 0$.

%%%%%%%%%%%%%%%%
%%%%%%%%%%%%%%%%

The particle production rate with such a spike-like mass is calculated using the result shown in App.~\ref{sec:spike}.
The axion energy density $\rho_a$ after one half saxion oscillation is given by
\begin{align}
	\frac{\rho_a}{\rho_\phi} \sim \frac{m_\phi^2 \widetilde \varphi_{\rm amp}^2}{f^4},
\end{align}
where $\rho_\phi \sim m_\phi^2 \widetilde \varphi_{\rm amp}^2$ is the saxion energy density.
Since we assumed $m_\phi \widetilde \varphi_{\rm amp} < f^2$, this ratio is smaller than one.
Formally we may define a ``decay rate'' of saxion into the axions as
\begin{align}
	\Gamma_{\varphi\to a} \simeq\frac{m_\phi \rho_a}{\rho_\phi} \sim \frac{m_\phi^3}{f^2}
	\left( \frac{\widetilde \varphi_{\rm amp}}{f} \right)^{2}.
\end{align}
It is clear from this expression that the axion production is significantly enhanced for $\widetilde \varphi_{\rm amp} \gg f$.
Moreover, one should take account of the effect of parametric resonance,
since the occupation number $f_k$ of the axion exceeds unity for $k \lesssim m_{\rm sp} \sim m_\phi \widetilde\varphi_{\rm amp}/f$.

We numerically evaluated the time evolution of the phase space density of the axion.
The linearized equation of motion of $\widetilde{a}_\phi$ in the momentum space is given by
\begin{align}
	\ddot {\widetilde a}_k + \omega_k^2(t)\widetilde a_k = 0,~~~~~~\omega_k^2(t) = k^2 + m_a^2(t),
\end{align}
where $m_a^2(t)$ is given by (\ref{ma2}) which is evaluated 
by numerically solving the equation of motion of saxion~(\ref{eom_saxion}).
The initial condition is taken to be 
${\widetilde a}_k = 1/\sqrt{2\omega_k}$ and $\dot{\widetilde a}_k = -i\sqrt{\omega_k/2}$.\footnote{
	This is the initial condition for the zero-point fluctuations in the Minkowski vacuum.
	Note that the long wave de-Sitter fluctuation generated during inflation may affect this initial condition depending on the inflation scale.
	However, the following result that the most enhanced mode is $k\sim m_\phi$ remains intact and 
	the phenomenological consequence also remains the same. 
}
The phase space density is given by
\begin{align}
	f_k^{(a)} = \frac{1}{2 \omega_k }\left( |\dot {\widetilde a}_k|^2 + \omega_k^2|{\widetilde a}_k|^2 \right) - \frac{1}{2}.
\end{align}
The right panel of Fig.~\ref{fig:fk_axion} shows the phase space density of the axion $f_k^{(a)}$ as a function of wave number $k$ in units of $m_\phi$
for 1, 3 and 5 half saxion oscillation $(t=\pi/m_\phi, 3\pi/m_\phi, 5\pi/m_\phi)$ for $\varphi_{\rm amp} = 100f$.
Hence the spike scale is $m_{\rm sp} \sim 100 m_\phi$.
It is seen that after one half saxion oscillation, the spectrum shows a $k^{-2}$ behavior for $m_\phi < k < m_{\rm sp}$
and cutoff for $k > m_{\rm sp}$ consistent with those studied in Refs.~\cite{Amin:2015ftc,Ema:2016dny}.
At this stage, the energy density is dominated by the modes $k \sim m_{\rm sp}$.
However, after a few oscillation, modes with $k \sim m_\phi$ is most enhanced and the axion energy density will soon be
comparable to the saxion energy density.
Then the backreaction will become important and the linearized treatment breaks down.
Another important feature is that the low frequency modes $k \ll m_\phi$ do not grow after a few oscillation
despite the fact that the occupation number is much larger than one.\footnote{
	Thus the long wave axion isocurvature fluctuation is not affected by this resonant enhancement.
}
This feature is much different from the ordinary broad resonance.
To understand this, one may find that Eq.~(\ref{eom_theta}) has a solution $\theta = {\rm const.}$ in the limit $k\to 0$,
due to the shift symmetry of $\theta$.
Actually we numerically checked that $\widetilde{a}_k(t) \propto \varphi(t)$ roughly followed for $k \ll m_\phi$.

It is difficult to rigorously follow the non-linear evolution of the whole system,
but it is reasonably expected that the axion backreacts to the saxion condensate so that 
the most saxion energy density is transferred to $k\sim m_\phi$ modes.
Then the saxion and axion will have comparable energy density until the mean saxion amplitude decreases to $\sim f$
due to the Hubble expansion.
After that, the axion becomes relativistic (see Eq.\,(\ref{ma})) and the axion energy density decreases faster than the saxion.
Since we are considering the initial saxion amplitude as large as the Planck scale, 
it is likely that the saxion dominates the universe at this stage.
Eventually the saxion perturbatively decays into MSSM particles: it decays into gluons in the KSVZ model~\cite{Kim:1979if} and into Higgs boson or higgsinos in the DFSZ model~\cite{Dine:1981rt}.
(Note that it does not perturbatively decay into the axion pair since we are assuming $\Delta m^2=0$.)
We parametrize the perturbative saxion decay rate as\footnote{
	The relation $\Gamma_\varphi / H( \widetilde\varphi_{\rm amp}=f) < 1$ is always satisfied
	as long as $m_\phi \widetilde\varphi_{\rm ini} < f^2$ and $m_\phi < f$.
}
\begin{align}
	\Gamma_\varphi = \frac{C}{8\pi} \frac{\widetilde{m}_\phi^3}{f^2},
\end{align}
where $C \sim \mathcal O(1)$ in the DFSZ model and $C\sim \mathcal O(10^{-3})$ in the KSVZ model.
The final axion radiation energy density in terms of the effective number of neutrino species is estimated as
\begin{align}
	\Delta N_{\rm eff} &\simeq \frac{43}{7}\left( \frac{10.75}{g_*(H=\Gamma_\phi)} \right)^{1/3}
	\left( \frac{C\widetilde{m}_\phi^2 M_P}{8\pi f^3} \right)^{2/3}\\
	& \sim 5\times 10^{-5} C^{2/3} \left( \frac{\widetilde{m}_\phi}{10^6\,{\rm GeV}} \right)^{4/3}
	\left( \frac{10^{12}\,{\rm GeV}}{f} \right)^{2},
	\label{Neff}
\end{align}
where $g_*$ denotes the relativistic degrees of freedom at the saxion decay.
This is a lower-bound on the axion dark radiation abundance in the sense that this amount is necessarily produced by the saxion oscillation
even if $\Delta m^2=0$ so that the perturbative saxion decay into the axion pair is forbidden.
Although $\Delta N_{\rm eff}$ is typically much smaller than the upper bound from the Planck observation~\cite{Ade:2015xua}, 
still it can have phenomenological impacts~\cite{Conlon:2013isa,Conlon:2013txa,Cicoli:2014bfa,Higaki:2013qka,Evoli:2016zhj}.
Gravitational waves produced due to the explosive axion production 
may be another interesting observable as discussed in Sec.~\ref{sec:dis}.

%%%%%%%%%%%%%%%%%%%%%%%%%%%%%%%%%%%%%%%%%%%%%%%%%%
\subsection{Axino}
%%%%%%%%%%%%%%%%%%%%%%%%%%%%%%%%%%%%%%%%%%%%%%%%%%

Next we study the production of the axino, or the fermionic superpartner of the axion.
There are three chiral fermions in the model (\ref{super}).
Denoting the fermionic components of $\phi, \bar\phi$ and $X$ by $\psi, \bar\psi$ and $\chi$, 
$\chi$ and one linear combination of $\psi$ and $\bar\psi$ obtain the Dirac mass of $\lambda\sqrt{|\phi|^2+|\bar\phi|^2}$.
On the other hand, there is a light mode, which we call the axino, defined by
\begin{align}
	A =\frac{1}{\sqrt{ |\phi|^2 + |\bar\phi|^2 }}\left( -\phi \psi + \bar\phi \bar\psi \right).
\end{align}
Its mass is given by
\begin{align}
	m_A = \lambda X = -\frac{4m_{3/2} f^2\varphi^2}{\varphi^4 + 4f^4}.  \label{m_axino}
\end{align}
Thus during the saxion oscillation the axino mass also oscillates and the axino is produced.
Below we evaluate the axino production rate for the small and large saxion amplitude regime, respectively.

%%%%%%%%%%%%%%%%%%%%%%%%%%%%%%%%%%%%%%%%%%%%%%%%%%
\subsubsection{Small saxion amplitude regime}  \label{sec:axino_small}
%%%%%%%%%%%%%%%%%%%%%%%%%%%%%%%%%%%%%%%%%%%%%%%%%%

First let us consider the axino production in the small saxion amplitude case: $\widetilde\varphi_{\rm amp} \ll f$.
The axino mass is expanded as
\begin{align}
	m_A \simeq - \frac{4 m_{3/2} f^2}{F^2 \left<\varphi\right>^2}
	+\frac{8 m_{3/2} f^2}{F^4 \left<\varphi\right>^3} \left(1 - \frac{4f^4}{\left<\varphi\right>^4} \right)\delta\varphi.
\end{align}
Thus the perturbative decay rate of the saxion into the axino pair is given by
\begin{align}
	\Gamma_{\varphi \to A} \simeq \frac{4 f^4 m_{3/2}^2 \widetilde m_{\phi}}{\pi F^{10} \left<\varphi\right>^6 }
	\left( \frac{\Delta m^2}{m_{\bar\phi}^2} \right)^2
	\left(1- \frac{4m_A^2}{\widetilde m_\phi^2} \right)^{3/2}.
\end{align}
Note that it also vanishes in the case of $\Delta m^2=0$.
Also such a process can be kinematically forbidden if $\widetilde m_\phi < 2 m_A$ even if $\Delta m^2\neq 0$.\footnote{
	Here we omitted the $\delta\varphi^2$ term in the expansion of the axino mass.
	Such a term induces the saxion ``annihilation'' into the axino pair even for $\Delta m^2=0$, although
	it is also kinematically blocked for $\widetilde m_\phi < m_A$.
}
Therefore it is possible to avoid the nonthermal axino production from the saxion dynamics as long as its amplitude is small enough.
However, as we shall see below, the axino production is unavoidable in the opposite regime $\widetilde\varphi_{\rm amp} \gg f$.
Below we assume $\Delta m^2=0$ to derive the lower bound on the axino abundance.

%%%%%%%%%%%%%%%%%%%%%%%%%%%%%%%%%%%%%%%%%%%%%%%%%%
\subsubsection{Large saxion amplitude regime}
%%%%%%%%%%%%%%%%%%%%%%%%%%%%%%%%%%%%%%%%%%%%%%%%%%

\begin{figure}[t]
\begin{center}
\begin{tabular}{cc}
\includegraphics[scale=1.2]{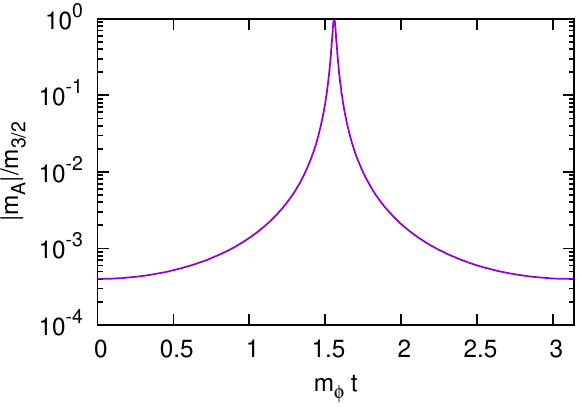}
\hspace{5mm}
\includegraphics[scale=1.2]{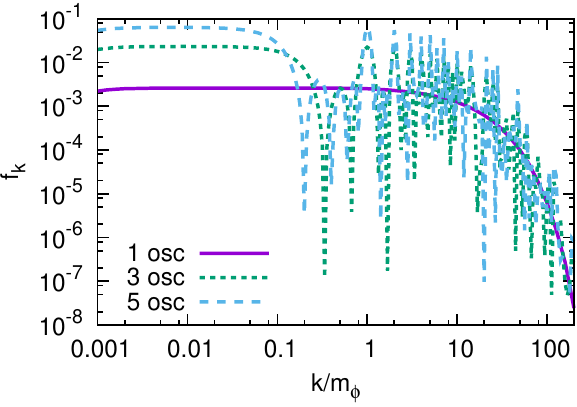}
\end{tabular}
\end{center}
\caption {
(Left) Time evolution of $|m_A(t)|$ for $\varphi_{\rm amp} = 100f$.
(Right) The phase space density of the axino $f_k$ as a function of wave number $k$ in units of $m_\phi$
for 1, 3 and 5 half saxion oscillation $(t=\pi/m_\phi, 3\pi/m_\phi, 5\pi/m_\phi)$ for $\varphi_{\rm amp} = 100f$.
}
\label{fig:fk_axino}
\end{figure}

Now we consider the large saxion regime: $\widetilde\varphi_{\rm amp} \gg f$.
In this case, the axino mass (\ref{m_axino}) shows a sharp behavior due to the saxon dynamics. 
Approximately we have
\begin{align}
	|m_A| \sim \begin{cases}
	\displaystyle \frac{4m_{3/2}f^2}{\widetilde\varphi^2} & {\rm for}~~~\left\lvert \widetilde\varphi \right\rvert \gg f,\\
	\displaystyle m_{3/2} & {\rm for}~~~\left\lvert \widetilde \varphi \right\rvert \ll f.
	\end{cases}
\end{align}
Fig.~\ref{fig:fk_axino} plots $|m_A|$ during one-half saxion oscillation for $\varphi_{\rm amp} = 100f$.
It is not hard to imagine that this peculiar behavior of the axino mass results in the axino production.

We numerically evaluated the time evolution of the phase space density of the axino.
The linearized equation of motion of the axino in the momentum space is given by
(see Refs.~\cite{Greene:1998nh,Peloso:2000hy,Asaka:2010kv,Ema:2016oxl} for more details on the fermion production)
\begin{align}
	\ddot {A}_k + \left(\omega_k^2(t) + i \dot m_A \right) A_k = 0,~~~~~~\omega_k^2(t) = k^2 + m_A^2(t),
\end{align}
with the initial condition $A_k =\sqrt{(\omega_k + m_A) / \omega_k}$ and $\dot A_k = -i\omega_k A_k$,
corresponding to the zero-point fluctuation in the Minkowski vacuum.
The phase space density is given by
\begin{align}
	f_k^{(A)} = \frac{1}{2\omega_k}\left[ m_A + 2 {\rm Im}\left( A_k^* \dot A_k \right) \right] + \frac{1}{2}.
\end{align}
The right panel of Fig.~\ref{fig:fk_axino} shows the phase space density of the axino $f_k^{(A)}$ as a function of wave number $k$ in units of $m_\phi$
for 1, 3 and 5 half saxion oscillation $(t=\pi/m_\phi, 3\pi/m_\phi, 5\pi/m_\phi)$ for $\varphi_{\rm amp} = 100f$.
We have taken $m_{3/2} = m_\phi$.
We find that one-half saxion oscillation yields $f_k ^{(A)}\sim (m_{3/2} / m_{\rm sp})^2$ for $m_A^{\rm min} \lesssim k \lesssim m_{\rm sp}$
where $m_A^{\rm min} \equiv 4 m_{3/2} f^2 / \widetilde\varphi_{\rm amp}^2$.
The production is not as violent as the axion case studied in the previous subsection, but still there can be a significant axino production.

The dominant contribution to the axino abundance may come from those created at $\widetilde\varphi_{\rm amp} \sim f$,
after which the axino production will be suppressed as shown in Sec.~\ref{sec:axino_small}.
The typical momentum of the axino is $\sim m_\phi$
and the phase space density around $k\sim m_\phi$ is likely to be saturated as $f_k^{(A)} = 1$ around this epoch.
The final axino yield after the saxion decay is estimated as
\begin{align}
	Y_A &\equiv \frac{n_A}{s} \sim \frac{3}{4\pi^2}\left( \frac{C}{8\pi} \right)^{1/2}\left( \frac{90}{\pi^2 g_*} \right)^{1/4} \frac{m_\phi^{5/2} M_P^{1/2}}{f^3 }\nonumber\\
	&\simeq 1\times 10^{-14}\,C^{1/2}  \left( \frac{{m}_\phi}{10^6\,{\rm GeV}} \right)^{5/2}\left( \frac{10^{12}\,{\rm GeV}}{f} \right)^{3}.
\end{align}
This may not be negligible depending on the mass of the LSP,
although it is smaller than the contribution from the thermal axino production~\cite{Brandenburg:2004du,Strumia:2010aa,Chun:2011zd,Bae:2011jb}.

%%%%%%%%%%%%%%%%
%%%%%%%%%%%%%%%%

%%%%%%%%%%%%%%%%%%%%%%%%%%%%%%%%%%%%%%%%%%%%
\section{Summary and discussion}  \label{sec:dis}
%%%%%%%%%%%%%%%%%%%%%%%%%%%%%%%%%%%%%%%%%%%%

We revisited the saxion dynamics in a SUSY axion model (\ref{super}) with two PQ fields,
and studied the axion production due to the saxion oscillation.
We found a strong spike-like behavior in the canonical axion field during the course of saxion oscillation,
which induces explosive production of axions if the saxion oscillation amplitude is larger than the PQ scale.
The production is so efficient that the axion energy density becomes comparable to the initial saxion energy density.
The typical momentum of the saxion and axion is of order the saxion mass scale $m_\phi$, and 
the initial saxion homogeneous condensate is expected to be transferred to the modes with $k \sim m_\phi$.
At this stage, both the saxion and axion have effective masses of order $m_\phi$ hence they are semi-relativistic.
Then the saxion and axion will keep to have a comparable energy density until the effective saxion amplitude becomes smaller than the PQ scale
due to the Hubble expansion.
After that the axion becomes relativistic while the saxion is non-relativistic and hence the saxion begins to dominate over the axion.
Eventually the saxion decays in a perturbative way.
This final saxion decay process is analyzed perturbatively around the vacuum and it is possible to suppress the 
decay into the axion pair by choosing appropriate parameters.
But the non-perturbative axion production is unavoidable when the saxion oscillation amplitude is large,
which translates into a lower bound on the axion dark radiation abundance (\ref{Neff}).

Some comments are in order.
We focused only on the axion and axino production since they are almost model independent.
Actually, however, the saxion has model-dependent couplings to the visible sector.
For example, in the KSVZ model the saxion couples to vector-like (s)quarks and in the DFSZ model 
it couples to the Higgs bosons and higgsinos.
These coupled particles are also efficiently produced due to the saxion oscillation through the preheating process.
In an extreme case, it may be possible that these processes make high temperature thermal bath
and the saxion and axion are thermalized much before the saxion decays perturbatively, 
although the complete analysis is highly model-dependent.
In any case, the most important feature that the saxion-axion system becomes 
inhomogeneous shortly after the onset of saxion oscillation remains intact.

One of the possible phenomenological consequences is the production of the gravitational waves (GWs) due to the violent axion production
and the non-linear dynamics of saxion-axion system.
Assuming that the saxion and axion are non-linearly inhomogeneous and their gradient energy density with a physical wave number $k\sim m_\phi$
are comparable to the total energy density of the universe,
the energy of GWs emitted from the region $k^{-1}$ in one Hubble time is evaluated as $\Delta E_{\rm GW} \sim H^{-1}M_P^{-2}(d^3 Q/dt^3)^2$ where $Q\sim \widetilde\varphi^2/k^3$ is the quadrupole moment of such a region.
Noting that there is $(m_\phi/H)^3$ such regions in the Hubble volume,
the relative GW energy density to the total energy density is given by 
$\omega_{\rm GW}\equiv \rho_{\rm GW} / \rho_{\rm tot} 
\sim \left(m_\phi/H\right)^3 \left(\widetilde{\varphi}/M_P\right)^4$.
It is $\mathcal O(1)$ just after the onset of saxion oscillation for $\widetilde{\varphi}_\mathrm{ini} \sim M_P$.
Thus the GW production is efficient at the early stage after the oscillation,
but it will be eventually diluted by the saxion-induced entropy production after the resonant axion production stops.
Quite roughly, we estimate the present abundance of the stochastic GW background as
\begin{align}
	\Omega_{\rm GW} &\sim \Omega_{\rm rad}\beta
	\left( \frac{\widetilde{\varphi}_\mathrm{ini}}{M_P}\right)^{2}
	\left( \frac{a(H=m_\phi)}{a(H=\Gamma_\phi)} \right) \nonumber \\
	&\sim 3\times 10^{-14} C^{2/3} \left( \frac{m_\phi}{10^6\,{\rm GeV}} \right)^{4/3}
	\left( \frac{10^{12}\,{\rm GeV}}{f} \right)^{4/3}
	\left( \frac{\widetilde{\varphi}_\mathrm{ini}}{M_P}\right)^{4/3},
\end{align}
where $\Omega_{\rm rad} \simeq 8.5\times 10^{-5}$ denotes the present radiation density parameter
and $\beta \simeq 0.3$ accounts for the deviation from $\rho_{\rm rad} \propto a^{-4}$ scaling law due to the change of relativistic degrees of freedom.
Note that it may overestimate the GW abundance by an order of magnitude.
The peak GW frequency at present in the present universe is
\begin{align}
	f_{\rm GW} &\sim \frac{m_\phi}{2\pi}\frac{a(H=m_\phi)}{a_0} \nonumber \\
	&\sim 50\,{\rm Hz}\,C^{1/6}\left( \frac{228.75}{g_*(H=\Gamma_\phi)} \right)^{1/3}
	\left( \frac{m_\phi}{10^6\,{\rm GeV}} \right)^{5/6}
	\left( \frac{10^{12}\,{\rm GeV}}{f} \right)^{1/3}
	\left( \frac{M_P}{\widetilde{\varphi}_\mathrm{ini}}\right)^{2/3}.
\end{align}
It may lie in the observable range in the future GW detectors such as DECIGO~\cite{Seto:2001qf}
for some reasonable parameter choices.
Note that typical physical wavenumber of GWs just after the emission is always $\sim m_\phi$,
hence GWs emitted later experience less redshift and constitute higher frequency modes in the present epoch.
The precise estimation of the GW spectrum requires numerical simulation, which is beyond the scope of this paper.

One may also wonder whether topological defects are formed or not due to the efficient axion production.
This issue is discussed in App.~\ref{sec:DW} where it is shown that we have no topological defects in this class of models.

%%%%%%%%%%%%%%%%%%%%%%%%%%%%%%%%%%%%%%%%%%%%
\section*{Acknowledgments}
%%%%%%%%%%%%%%%%%%%%%%%%%%%%%%%%%%%%%%%%%%%%

K.N. would like to thank K.~Saikawa for useful discussion.
The work of Y.E. was supported in part by JSPS Research Fellowships for Young Scientists.
The work of Y.E. was also supported in part by the Program for Leading Graduate Schools, MEXT, Japan.
This work was supported by the Grant-in-Aid for Scientific Research on Scientific Research A (No.26247042 [KN]),
Young Scientists B (No.26800121 [KN]) and Innovative Areas (No.26104009 [KN], No.15H05888 [KN], No.17H01131 [KN]).

\appendix
%%%%%%%%%%%%%%%%%%%%%%%%%%%%%%%%%%%%%%%%%%%%
\section{Particle production with spike mass}  \label{sec:spike}
%%%%%%%%%%%%%%%%%%%%%%%%%%%%%%%%%%%%%%%%%%%%

Let us consider the production of scalar $\chi$ particle under the time dependent mass term $m_\chi^2(t)$.
We parameterize the spike-like time-dependent mass term as
\begin{align}
	m_\chi^2(t) = \frac{\mu^2}{2}{\rm sech}^2(m_{\rm sp}(t-t_0)) + m_{\chi 0}^2,
\end{align}
where $m_{\chi 0}^2 < m_{\rm sp}^2$ is assumed.
Such a situation was analyzed in App.~C of Ref.~\cite{Amin:2015ftc} (see also App.~B of Ref.~\cite{Ema:2016dny}).
For $t\gg t_0$, the phase space distribution of $\chi$ particle is given by
\begin{align}
	f^{(\chi)}_k \simeq \mathcal C \times \begin{cases}
		\left[m_{\rm sp}/(\pi m_{\chi 0}) \right]^2 & {\rm for}~~~k \ll m_{\chi 0},\\
		\left[m_{\rm sp}/(\pi k) \right]^2   & {\rm for}~~~m_{\chi 0}\ll k \ll m_{\rm sp},\\
		\exp\left(-2\pi k/m_{\rm sp}\right) & {\rm for}~~~k \gg m_{\rm sp},
	\end{cases}
\end{align}
where
\begin{align}
	\mathcal C \simeq 
	\begin{cases}
		(\mu/m_{\rm sp})^4   & {\rm for}~~~\mu \ll m_{\rm sp},\\
		1 & {\rm for}~~~\mu \gg m_{\rm sp}.
	\end{cases}
\end{align}
Thus the energy density of produced $\chi$ particles is dominated by those with momentum $k\sim m_{\rm sp}$ and
it is given by
\begin{align}
	\rho_\chi \sim  
	\begin{cases}
		\mu^4   & {\rm for}~~~\mu \ll m_{\rm sp},\\
		 m_{\rm sp}^4 & {\rm for}~~~\mu \gg m_{\rm sp}.
	\end{cases}
\end{align}

In the case of axion production studied in the main text, we have
\begin{align}
	\mu \simeq m_\phi\left(\frac{\widetilde\varphi_{\rm amp}}{f}\right),~~~~~m_{\rm sp} \simeq m_\phi\left(\frac{\widetilde\varphi_{\rm amp}}{f}\right),
	~~~~~m_{\chi 0} \simeq m_\phi.
\end{align}
It is consistent with numerical calculation shown in Fig.~\ref{fig:fk_axion}.
Thus particles with $f_k \gtrsim 1$ $(k < m_{\rm sp})$ are produced after one saxion oscillation.
After that, the production is even more enhanced through the parametric resonance or the Bose enhancement effects.

%%%%%%%%%%%%%%%%%%%%%%%%%%%%%%%%%%%%%%%%%%%%
\section{Comment on domain wall formation}  \label{sec:DW}
%%%%%%%%%%%%%%%%%%%%%%%%%%%%%%%%%%%%%%%%%%%%

%%%%%%%%%%%%%%%%
\begin{figure}[t]
\begin{center}
\begin{tabular}{cc}
\includegraphics[scale=1.2]{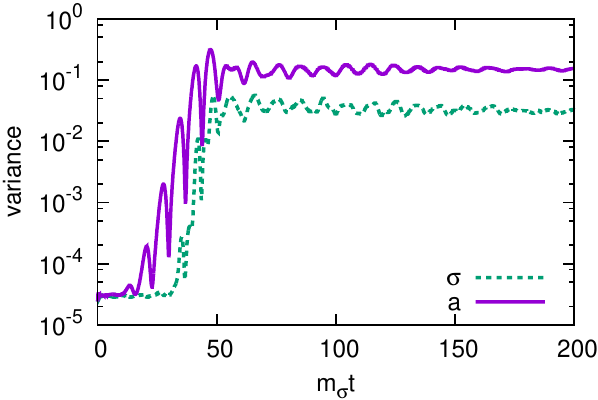}
\hspace{3mm}
\includegraphics[scale=1.2]{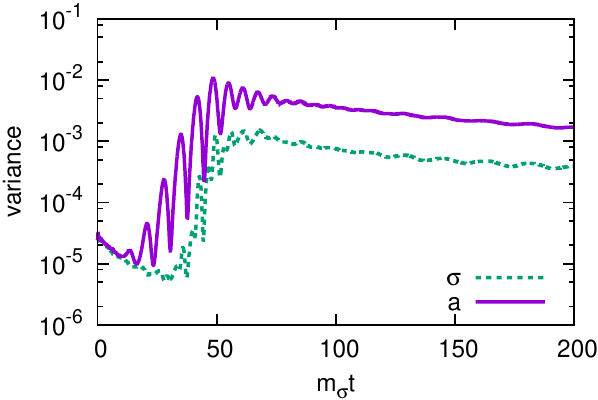}
\end{tabular}
\end{center}
\caption {
The evolution of the field variance $\left< a^2 \right> - \left< a \right>^2$ and $\left< \sigma^2 \right> - \left< \sigma \right>^2$
(in units of $f^2$) are shown as a function of time $m_{\sigma}t$.
The initial condition is set to be $\sigma_i = 0.3 f$ and we have taken $\lambda = 10^{-5}$ and $f = 0.1 M_P$.
(Left) The Hubble expansion is not included.
(Right) Power-law Hubble expansion is assumed.
}
\label{fig:lattice}
\end{figure}
%%%%%%%%%%%%%%%%

Here we comment on the possibility of domain wall (DW) formation due to the saxion oscillation.
It is known that if the PQ field oscillates with large amplitude, the parametric resonant enhancement of the axion fluctuation
can lead to the nonthermal symmetry restoration~\cite{Kofman:1995fi,Kasuya:1996ns,Kasuya:1997ha,Kawasaki:2013iha}.
In such a case, the formation of topological defects, i.e., cosmic strings after the PQ phase transition and DWs after the QCD phase transition, would be unavoidable.
 
However, Ref.~\cite{Kasuya:1996ns} also claimed that DWs may be formed even in the case without nonthermal symmetry restoration,
because the saxion oscillation in a SUSY axion model leads to the resonant enhancement of the axion fluctuation
and the phase of the PQ field might be uniformly distributed in the universe.
Below we show that the DW formation does not likely to happen in such a setup.
Let us consider a toy model
\begin{align}
	\mathcal L = - \left|\partial \phi \right|^2 - \lambda \left( |\phi|^2 - \frac{f^2}{2} \right)^2.
\end{align}
Writing $\phi = (f + \sigma + ia)/\sqrt{2}$, we set the initial condition $\left<\sigma\right>=\sigma_i$ and $\left<a\right>=0$.
If $0< \sigma_i < (\sqrt 2-1)f$, the initial potential energy is lower than the potential energy at the origin $\phi=0$
and hence nonthermal symmetry restoration cannot occur with this initial condition.\footnote{
	One can consider a model in which $\phi$ has a flat potential for $|\phi| \gtrsim f$. 
	In such a case, this kind of initial condition is naturally realized~\cite{Ema:2016ops}.
}
Still, the axion fluctuation as well as the fluctuation of the radial mode is amplified during the $\sigma$ oscillation around the vacuum $\sigma=0$.
The question is whether such an enhancement of the axion fluctuation leads to the formation of DWs or not.
A naive estimation is that the parametric resonance stops when the axion fluctuation $\delta a^2\equiv \left< a^2 \right> - \left< a \right>^2$
becomes roughly equal to $f^2$, around which the backreaction becomes important.
Thus we need to simulate it numerically to correctly estimate the field variance.

We performed classical lattice simulation using {\tt LATTICEEASY}~\cite{Felder:2000hq}.
The initial condition is set to be $\sigma_i = 0.3 f$ and we have taken $\lambda = 10^{-5}$ and $f = 0.1 M_P$
with the grid size of $128^3$.
The results are shown in Fig.~\ref{fig:lattice}.
The evolution of the field variance $\delta a^2$ and $\delta\sigma^2\equiv \left< \sigma^2 \right> - \left< \sigma \right>^2$
(in units of $f^2$) are shown as a function of time $m_{\sigma}t$ where $m_{\sigma} = \sqrt{2\lambda} f$.
The Hubble expansion is not included in the left figure while
a power-law Hubble expansion $R(t) \propto t^{1/2}$ is assumed in the right figure.
It is seen that, although the fluctuation is resonantly enhanced, the resonance stops before reaching $\delta a^2 \simeq f^2$.
It means that the axion field is not uniformly distributed from $\theta= 0$ to $\pm\pi$ where $\theta \equiv {\rm arg}(\phi)$
even after the parametric resonance.
Moreover, the fluctuation decreases as the universe expands.\footnote{
	This is because the resonantly enhanced axion modes are sub-horizon.
	If (extremely) superhorizon modes would be enhanced, it can have impacts on DW formation but 
	we checked that this is not the case as mentioned in Sec.~\ref{sec:axion}.
}
Since the QCD phase transition happens far after the $\sigma$ field oscillation, we expect $\theta$ is almost zero in
the whole universe around the QCD phase transition epoch, and hence there is no DW formation.
The story is the same for the SUSY axion model studied in the main text.

%%%%%%%%%%%%%%%%%%%%%%%%%%%%%%%%%%%%%%%%%%%%%%%%%%

%%%%%%%%%%%%%%%%%%%%%%%%%%%%%%%%%%%%%%%%%%%%%%%%%%

\end{document}